# Thickness-Controlled Ferroelectric-to-Ionic Transport in $CuCrP_2S_6$ at Room Temperature

Subhashree Chatterjee, Abhishek Bajgain, Rabindra Basnet, Ramesh C. Budhani*

Department of Physics, Morgan State University, Baltimore, MD, 21251, USA

*E-mail: ramesh.budhani@morgan.edu

Ambient-temperature coexistence of ferroelectricity, photoconductivity, and ionic transport in ultra-thin van der Waals (vdW) compound $CuCrP_2S_6$ (CCPS) offers significant potential for multifunctional electronic, photonic, and neuromorphic applications. However, ferroelectricity and ionic transport can become antagonistic when sharing a common atomic origin, like the copper ions in CCPS. We address this fundamental dichotomy by systematically investigating the thickness-dependent evolution of ferroelectricity, ionic migration, and memristive behavior in CCPS diodes via electrostatic and photo-assisted interfacial band tuning. Using piezoresponse, Kelvin probe, and conductive atomic force microscopy, alongside photoexcited transport measurements, we identify distinct thickness regimes. CCPS flakes thinner than ~ 40 nm exhibit robust, reversible ferroelectric (FE) switching. Conversely, thicker samples display suppressed switchable polarization, with electrical conduction dominated by $Cu^+$ migration, filamentary memristive switching, and rectification reversal. Optical illumination further enhances ionic mobility via photogenerated carriers, driving photo-assisted ionic transport in non-FE flakes and a coupled photoferroionic response in FE flakes. These findings clarify how ferroelectricity fundamentally alters ionic transport dynamics in vdW ferroionic systems.

# 1. Introduction

The discovery of ferroelectricity in two-dimensional van der Waals (vdW) materials has opened new opportunities for realizing ultrathin nonvolatile memories[1,2], neuromorphic systems[3], and reconfigurable optoelectronic devices[4]. Unlike conventional oxide ferroelectrics (FEs), many vdW FEs derive their polarization from the displacement of weakly bonded ions within layered crystal structures[5], leading to a coupling between FE order and ionic dynamics[6–8]. This coexistence of switchable polarization and mobile ionic species gives rise to ferroionic behavior[7,9,10], where polarization switching, charge transport, and ionic migration are strongly coupled. Understanding the competition and cooperation among these processes is therefore essential for the development of next-generation multifunctional devices[1,6,11–14]. Among vdW ferroionic materials, $CuCrP_2S_6$ (CCPS) has emerged as a particularly attractive platform owing to its coupled FE, ionic, semiconducting, and magnetic properties[15–18,19]. Its layered crystal structure consists of a weakly bonded sulfur-phosphorus framework hosting mobile $Cu^+$ ions and magnetic Cr ions[15–17]. The displacement and ordering of $Cu^+$ ions break inversion symmetry and generate electric polarization[7,20,21], while their relatively low migration barrier enables ionic transport under external electric fields[7,22–27]. As a result, CCPS represents a rare material system in which ferroelectricity and ion migration originate from the same microscopic degree of freedom. This duality enables unconventional transport phenomena including memristive switching[28], ferroionic conduction[29], and electrostatic modulation of transport[30,31].

Bulk CCPS exhibits antiferromagnetic ordering of $Cr^{3+}$ ions below the Néel temperature ($T_N \approx 32$ K)[18] along with strongly anisotropic electrical and magnetic responses[17]. The FE state emerges from the temperature-dependent ordering of $Cu^+$ ions within octahedral sulfur cages[28]. However, dipoles in alternate Cu-containing planes of the vdW structure are ordered antiferroelectrically and the system transitions to a paraelectric phase above $T_C$ ~190 K[32].

Intriguingly, this bulk limitation appears to be altered in reduced dimensions, where recent studies have reported room-temperature FE behavior in ultrathin CCPS flakes[28,33], with an intrinsic spontaneous polarization ~ 16.05 μC $cm^{-2}$ [33]. At the same time, $Cu^+$ ion migration has been shown to strongly influence current (*I*)–voltage (*V*) characteristics and surface electrostatics in two-terminal devices of CCPS[7,25,28]. Such ionic motion can also produce hysteresis and memristive behavior that closely resembles FE switching, making it difficult to clearly distinguish between ionic and polarization-driven contributions to charge transport. Accordingly, the neuromorphic synaptic behavior observed in *I-V* characteristics of CCPS devices has either been attributed to ionic migration[34] or to FE polarization modulation[28]. Despite these advances, a unified understanding of how FE ordering and $Cu^+$ ion migration compete or cooperate, and how their relative contributions evolve with thickness remains lacking. In addition, CCPS exhibits strong photoresponse in its transport characteristics[35]. Photogenerated carriers can alter local electrostatic potentials, thereby influencing both polarization stability[36,37] and $Cu^+$ ion migration through field-assisted ionic motion, similar to other Cu-based vdW ferroionic systems such as $CuInP_2S_6$[35,36]. These intertwined couplings between dimensionality, polarization, ionic transport, and photoexcitation motivate a systematic investigation of their combined influence on charge transport in CCPS nanoflakes to realize functional photoresponsive ferroionic devices.

Here, we systematically investigate the thickness-dependent evolution of ferroelectricity, surface electrostatics, ionic migration, and charge transport mechanisms in CCPS nanoflakes by combining piezoresponse force microscopy (PFM), Kelvin probe force microscopy (KPFM), sweep-rate-dependent conductive atomic force microscopy (C-AFM) under dark and illuminated conditions, and measurements of *I-V* characteristics of planar device. We identify a crossover from ferroelectricity-assisted ferroionic transport in ultrathin flakes to ionic-migration-dominated memristive conduction in thicker flakes. We further show that optical illumination enhances transport differently in the two regimes, giving rise to photoferroionic behavior in thinner flakes and photo-assisted ionic memristive transport in thicker flakes. These findings establish CCPS as a model vdW ferroionic system in which ferroelectricity actively governs ionic transport and memristive functionality.

## 2. Results and Discussion

The central objective of this work is to investigate the interplay between polarization-related effects and ionic transport in CCPS across different thickness regimes. By combining PFM, KPFM, C-AFM, and device-level electrical measurements, we explore the thickness-dependent evolution of ferroelectric response and charge transport behavior in CCPS nanoflakes. We further show that optical excitation modulates transport differently across different thickness regimes. In the following sections, we present measurements performed on well-characterized CCPS single crystals (Figure S1, Supporting Information). A schematic of the experimental configuration used for PFM, KPFM, and C-AFM measurements under controlled optical illumination is shown in Figure S2.

**Thickness-dependent room-temperature ferroelectricity in CCPS**

To investigate the nanoscale FE response, out-of-plane (OOP) PFM measurements were performed on flakes with varying thicknesses. Representative topography, PFM phase, and amplitude images of a CCPS crystal of thickness ~15 nm - 62 nm are shown in Figure S3 (Supporting Information). To probe the FE nature, local domain engineering was carried out using a box-in-box poling protocol. Figures 1a–h show PFM phase images acquired after application of reverse DC bias (±10 V), on different thickness of flakes (12 nm - 62 nm). The corresponding PFM amplitude images are shown in Figure S4 (Supporting Information). Clear and reversible phase contrast is observed in flakes with thickness ≤ 40 nm. In these regions, the written domain patterns remain stable even after a time window of 2 hours (Figures 1f,g), indicating non-volatile polarization-like behavior at room temperature within this thickness regime.

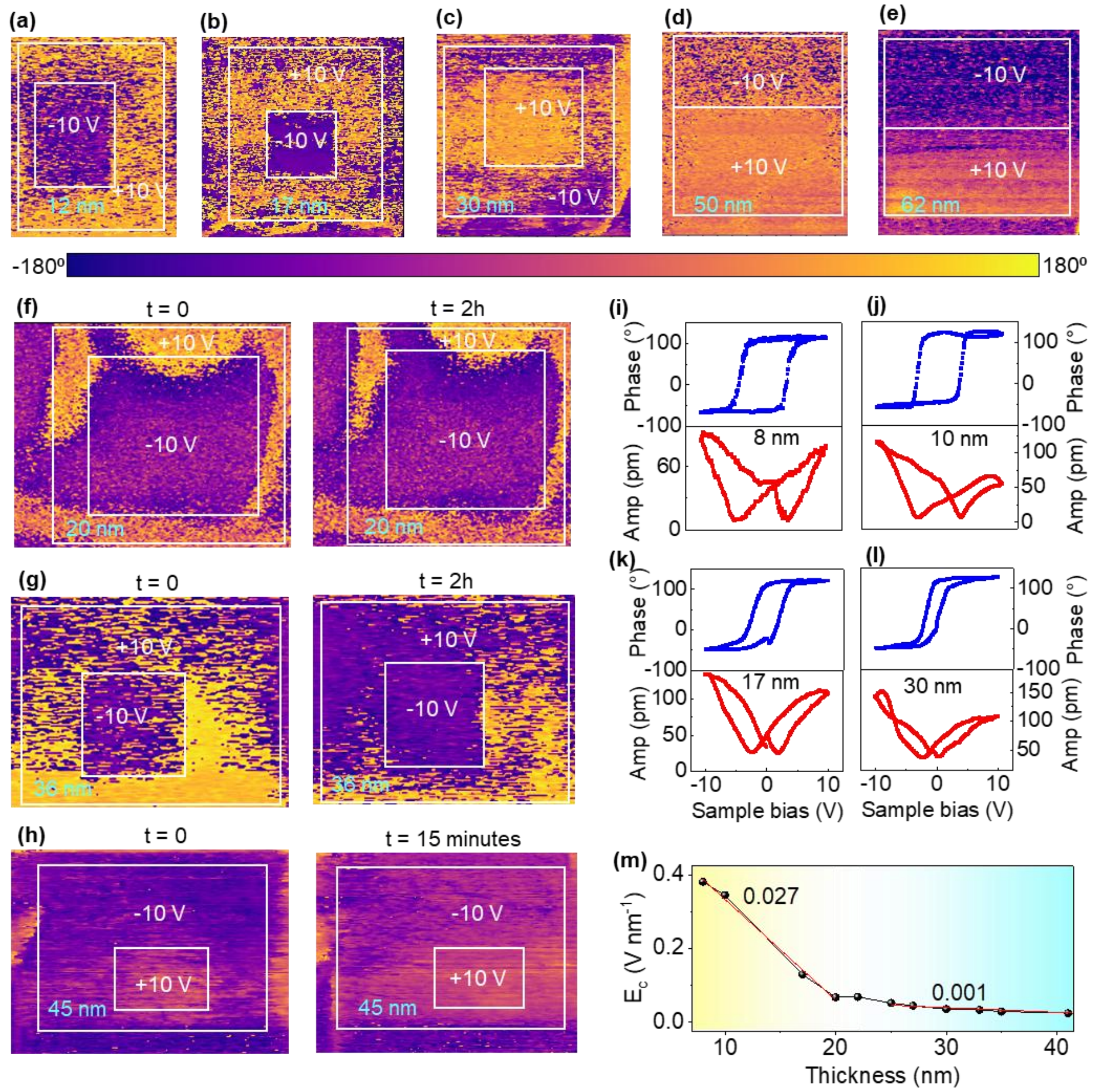

***Figure 1. Thickness-dependent room temperature OOP PFM measurements in CCPS nanoflakes.*** *a–e) OOP-PFM phase images of CCPS flakes with thicknesses of 12, 17, 30, 50, and 62 nm, respectively, following local domain writing using reversible DC poling under ±10 V bias. f–h) Time evolution of written domains in CCPS flakes with thicknesses of 20, 36, and 45 nm, showing PFM phase images acquired immediately after poling (t = 0) and after a*

*waiting time of 2 h (t = 2 h), indicating the temporal stability of the written state. i–l) OOP-PFM switching spectroscopy (off-state measurements) for CCPS flakes with thicknesses of 8, 10, 17, and 30 nm. The upper panels show phase hysteresis loops, while the lower panels present the corresponding amplitude (butterfly-shaped) responses, consistent with switchable polarization-like electromechanical behavior. m)* $E_c$ *as a function of flake thickness. Solid lines represent linear fits used to guide the eye and highlight the thickness-dependent trend.*

A systematic thickness-dependence is observed in the domain response. In very thin flakes (< 20 nm), the written domains exhibit reduced contrast and spatial non-uniformity, suggesting that weaker electrostatic screening and enhanced depolarization effects influence the stability of the switched state[38,39]. The intermediate thickness flakes (~20–36 nm) show comparatively more uniform domain patterns, indicating a stable switching behavior. For thicknesses above ~40 nm, the written contrast becomes progressively weaker, suggesting a reduced efficiency of local polarization switching under identical experimental conditions. The flake of thickness ~20 nm exhibits the highest stability in domain retention, with minimal decay in phase contrast over 2 hours, whereas thicker flakes show progressively faster relaxation of the written state. Earlier reports have shown that Cu-based vdW chalcogenides can exhibit bias-induced surface modification associated with $Cu^+$-ion redistribution[24,40] under prolonged electrical stress. In the present measurements, no noticeable surface topographic changes were observed after a single poling cycle under the applied experimental conditions, suggesting that irreversible electrochemical modification is not dominant within the measurement window. However, given the known mobility of $Cu^+$ ions in CCPS, their contribution to the tip-sample interaction cannot be completely excluded. To quantify the switching behavior, local PFM spectroscopy was performed on flakes of different thicknesses (Figures 1i–l). All measured flakes exhibit phase hysteresis between forward and reverse bias sweeps, together with characteristic butterfly-shaped amplitude responses, consistent with a switchable polarization-like response.

The switching characteristics show a clear dependence on thickness: ultrathin flakes (8–10 nm) exhibit broader hysteresis and larger effective switching voltage, while intermediate thickness flakes show comparatively sharper switching features. The extracted effective coercive field ($E_c$ = coercive voltage/thickness) shows two distinct regimes as a function of thickness (Figure 1m), as indicated by two linear fitting lines with different slopes. In the ultrathin regime, $E_c$ decreases rapidly with increasing thickness, while at larger thicknesses the dependence becomes weaker, suggesting a change in the dominant factors governing switching, such as surface effects, incomplete electrostatic screening, and enhanced interface pinning effects[33,37-39]. Given the ferroionic nature of CCPS, $Cu^+$ ion migration may also contribute to the observed electromechanical switching response by modifying local electrostatic conditions under applied bias. Therefore, while the observed PFM response, including reversible phase contrast, hysteresis-like switching, and FE-like retention is consistent with polarization reversal, it may be interpreted as arising from a coupled ferroionic system in which both polarization-related and ionic contributions may coexist. Within this framework, the ultrathin regime exhibits the strongest polarization-like contribution to the overall switching response under the present measurement conditions.

**Thickness-dependent surface potential modulation probed by KPFM**

To probe the evolution of the local electrostatic environment in CCPS, KPFM measurements were performed on flakes of varying thickness. Figures 2a,b show representative topography and surface potential ($V_{CPD} = (\phi_{tip} - \phi_{CCPS})/e$, where $\phi$ denotes work function, and e is the elementary charge)[37] maps for flakes with thicknesses ranging from ~5 to 80 nm. A clear thickness-dependence of $V_{CPD}$ is observed, as summarized in Figure 2c, exhibiting a non-monotonic behavior with a maximum in the intermediate thickness regime (~40 nm), while both thinner and thicker flakes show comparatively lower surface potential contrast.

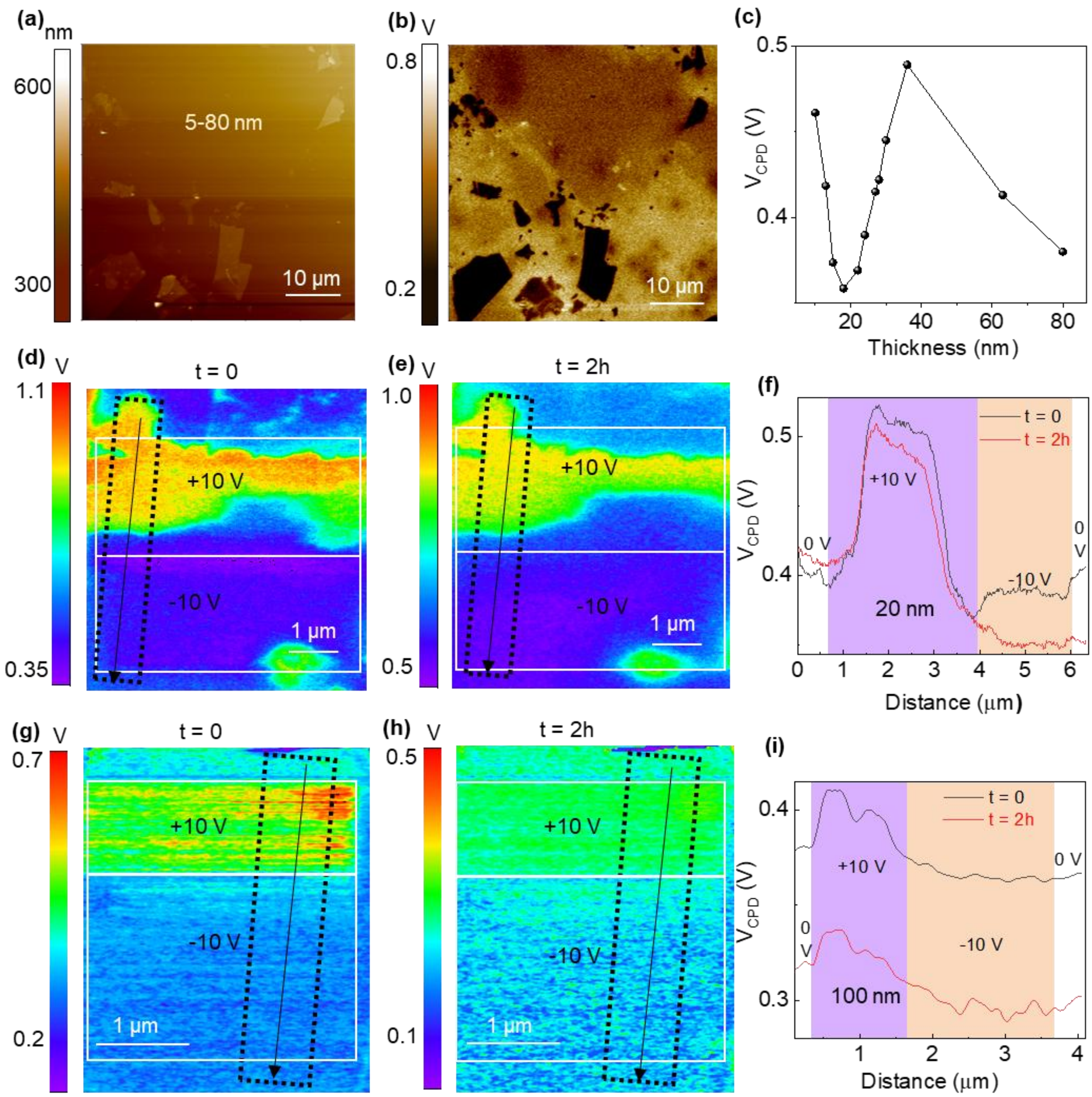

***Figure 2. Thickness-dependent surface potential modulation in CCPS probed by KPFM.*** *a) AFM topography image of CCPS flakes with varying thickness exfoliated on a conductive Si substrate. b) Corresponding KPFM surface potential* $V_{CPD}$ *map of the same flakes. c) Extracted* $V_{CPD}$ *as a function of flake thickness, showing a non-monotonic dependence with a maximum in the intermediate thickness regime. d,e)* $V_{CPD}$ *maps of a representative ~20 nm flake acquired after local poling with ±10 V DC bias d), and after a relaxation time of 2 h (**e**), showing evolution of the written electrostatic contrast. f) Line profiles of* $V_{CPD}$ *across the poled regions of the ~20 nm flake, extracted from the area marked by the black dashed box in d,e), shown*

*immediately after poling (t = 0, black) and after 2 h (red). g,h) $V_{CPD}$ maps of a thicker (~100 nm) CCPS flake measured immediately after local poling g) and after 2 h of relaxation h), showing a reduced persistence of the written contrast compared to thinner flakes. i) Corresponding $V_{CPD}$ line profiles across the poled regions of the ~100 nm flake, extracted from the black dashed regions in g,h), for t = 0 (black) and t = 2 h (red). The different color bands in f) and i) represent $V_{CPD}$ for spatially poled regions under +10 V and -10 V bias, respectively, measured along the black arrow lines.*

This thickness-dependent variation demonstrates that the local electrostatic landscape is strongly influenced by the combined effects of polarization-related charge distribution, interfacial screening, and ionic contributions in CCPS. To further examine local electrostatic modulation under external bias, controlled poling experiments were performed on a representative ~20 nm thick flake. Figures 2d,e show $V_{\mathrm{CPD}}$ maps acquired immediately after application of ±10 V DC bias and after a relaxation time of 2 h. A clear contrast between oppositely poled regions is observed immediately after writing, and a significant portion of this contrast persists after 2 h. The corresponding line profiles (Figure 2f) reveal a surface potential difference exceeding ~100 mV between the two written states in the initial measurement. The persistence of surface potential contrast over time indicates the presence of a relatively stable remnant electrostatic state, although contributions from slow charge relaxation and ionic redistribution cannot be completely excluded. Importantly, no measurable topographic changes were observed after the poling procedure, suggesting that the observed potential contrast is not accompanied by permanent surface deformation. For comparison, similar measurements were performed on a thicker (~100 nm) flake (Figures 2g–i). While an initial surface potential contrast is observed immediately after poling, the $V_{\mathrm{CPD}}$ difference significantly decreases after 2 h, and spatial contrast becomes less distinct compared to the thinner flake. This reduced retention suggests enhanced relaxation processes and weaker stability of the written electrostatic state in the thicker sample of CCPS. The thickness-dependent evolution of surface potential and retention observed here in KPFM is consistent with the trends identified in PFM measurements, indicating stronger polarization-related contributions in the ultrathin regime and a progressive increase in relaxation-driven and ionic contributions with increasing thickness. To investigate the implications of this crossover on charge transport, we next examine the results of C-AFM measurements.

**Transport mechanisms in ultrathin and thicker CCPS flakes**

As ionic migration occurs on significantly longer timescales than electronic transport, varying the voltage sweep-rate provides a useful means of probing the relative contributions of electronic and ionic processes to the C-AFM current response. Figures 3a,b show the $I$–$V$ characteristics measured at sweep-rates between 15 and 1 V $s^{-1}$ for the 110 nm and 20 nm flakes, respectively. The measurements were performed using two semi-cycles of a triangular voltage waveform (0 V → +9 V → 0 V and 0 V → −9 V → 0 V), while the complete voltage-cycle responses are presented in Figure S5. The two flakes exhibit strongly asymmetric conduction and pronounced hysteresis on bias reversal; however, their evolution with sweep-rate differs substantially. At high sweep-rates (15 – 10 V $s^{-1}$), both samples display rectifying transport behavior consistent with the asymmetric Si/CCPS/Pt device geometry. Comparison measurements performed on symmetric Pt/CCPS/Pt structures (Figure S6) further confirm the importance of interface asymmetry in determining the transport characteristics.

To quantify the voltage polarity-dependent conduction, the rectification ratio was defined as $R = |I_{-}/I_{+}|$, where $I_{-}$ and $I_{+}$ are the currents measured at the maximum negative and positive voltages, respectively. As summarized in Figure 3i, the ultrathin flake exhibits a substantially larger rectification ratio than the thick flake over the entire sweep-rate range.

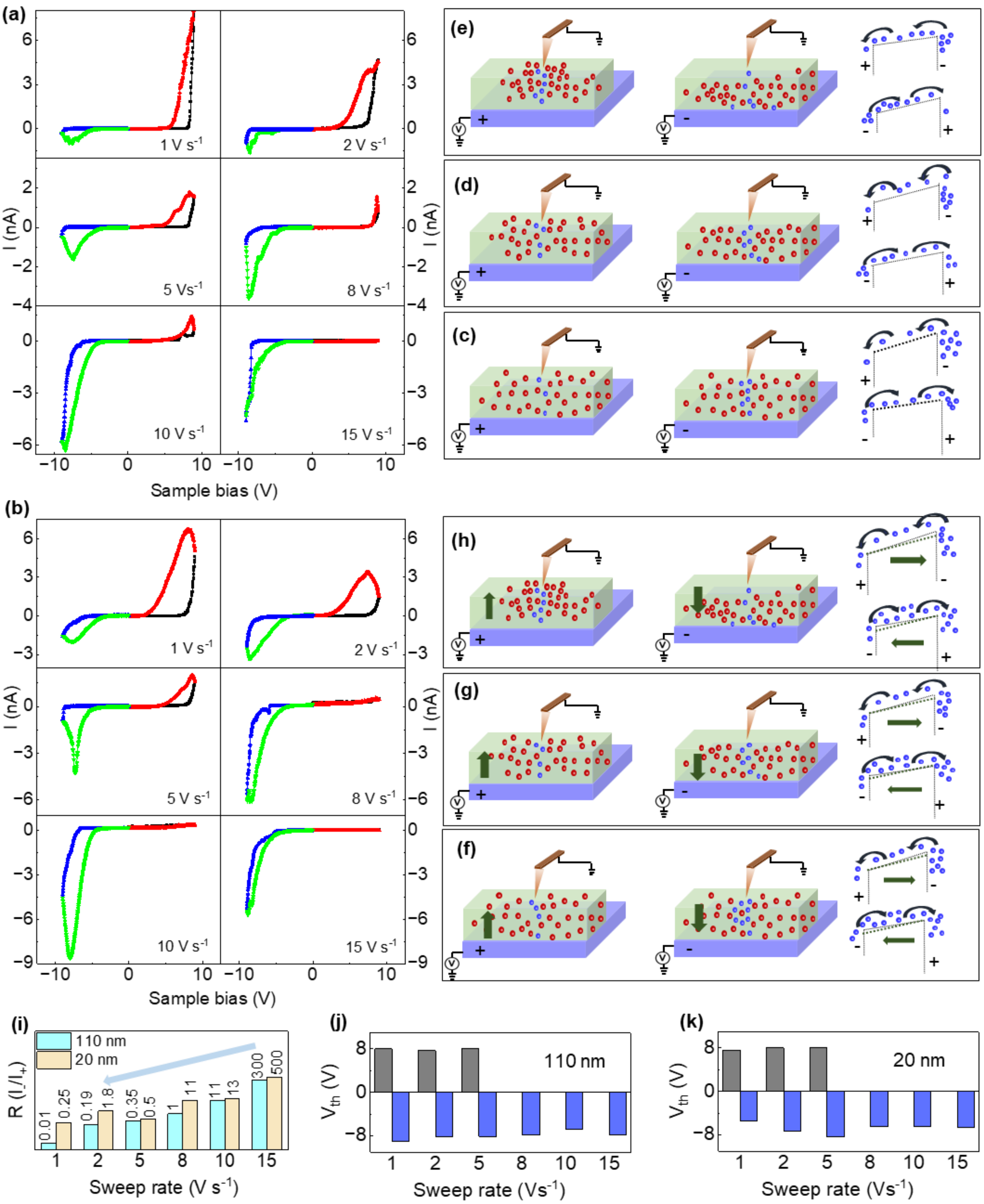

***Figure 3. Sweep-rate-dependent transport evolution in thick and ultrathin CCPS flakes.*** *a,b) C-AFM I–V characteristics measured at different voltage sweep-rates on representative thick*

*(~110 nm) and ultrathin (~20 nm) CCPS flakes, respectively, using an asymmetric Si/CCPS/Pt junction geometry. Measurements were acquired using two triangular voltage semi-cycles: 0 V → +9 V → 0 V (black and red traces) and 0 V → −9 V → 0 V (blue and green traces). c–e) Proposed transport scenarios for the thick (~110 nm) flake at high, intermediate, and low sweep-rates, respectively, illustrating the increasing influence of $Cu^+$ redistribution on interfacial band profiles and charge transport. Red and blue spheres represent the $Cu^+$ ions and electrons, respectively. f–h) Corresponding transport scenarios for the ultrathin (~20 nm) flake, highlighting the interplay between polarization-related electrostatic modulation and ionic redistribution. Green arrows denote the polarization directions. i) Rectification ratio ($R = |I_-/I_+|$) numbers as a function of sweep-rate for the thick and ultrathin flakes (vertical bars not to the scale). j,k) Evolution of the $V_{th}$ with sweep-rate for the thick (~110 nm) and ultrathin (~20 nm) CCPS flakes, respectively. Numbers along the x-axis in i-k) represent the value of voltage sweep-rates, not the scale. The schematics in c–h) are proposed transport models consistent with the experimental observations.*

Since both devices were measured using the same electrode configuration, this enhanced rectification suggests an additional electrostatic contribution in the ultrathin regime. In conjunction with the persistent polarization-induced surface-potential contrast observed by KPFM (Figure 2), the larger rectification is consistent with polarization-assisted modulation of the carrier injection barrier. The transport response of the thick (~110 nm) flake evolves strongly with decreasing sweep-rate. At fast voltage sweeps, the current remains relatively small under positive bias, whereas conduction is favored under negative bias owing to the asymmetric interfacial barriers (Figure 3c). As the sweep-rate is reduced, progressively larger positive-bias currents emerge, leading to a gradual reduction and eventual reversal of the rectification polarity ($R < 1$ in Figure 3i). This pronounced sweep-rate-dependence indicates the increasing participation of the slow field-driven processes in charge transport. Given the mobility of $Cu^+$ ions in CCPS, the observed behavior is consistent with electric-field-induced ionic redistribution, which modifies the local electrostatic environment near the tip–sample interface and facilitates carrier injection (Figures 3d,e). At the slowest sweep-rates, the current increases abruptly under positive bias, producing highly conductive states characteristic of ionic-migration-assisted memristive transport. The observed current evolution is consistent with the formation of localized conductive pathways[41] associated with $Cu^+$-ion accumulation. The overall trend in sweep-rate-dependence and rectification reversal indicate that ionic migration becomes the dominant transport mechanism in the thick-flake regime. A markedly different behavior is observed in the ultrathin (~20 nm) flake (Figure 3b). At high sweep-rates, current turn-on occurs at substantially lower voltages than in the thick flake, particularly under negative bias. The corresponding threshold voltages ($V_{th}$) extracted from the $I$–$V$ curves are summarized in Figures 3j,k. Whereas the thick flake exhibits $V_{th}$ close to ±8–9 V (Figure 3j) throughout the investigated sweep-rate range, the ultrathin flake consistently displays lower values, with $V_{th}$ decreasing to approximately −5.4 V at 1 V s$^{-1}$. This reduction in turn-on voltage is consistent with polarization-induced electrostatic modulation of the interfacial band profile, which lowers the effective carrier-injection barrier. Notably, the KPFM measurements reveal a polarization-dependent surface-potential contrast exceeding ~100 mV (Figure 2), corresponding to an estimated interfacial barrier modulation of the order of ~0.1 eV. Such electrostatic modulation provides a plausible origin for the reduced $V_{th}$ and enhanced rectification observed in the ultrathin flakes. As the sweep rate decreases below ~8 V s$^{-1}$, the hysteresis loops of the ultrathin flake evolve gradually, indicating increasing interaction between polarization-related electrostatic fields and ionic redistribution (Figures 3f–h). In contrast to the thick flake, however, no abrupt rectification reversal or sudden current jump is observed. Instead, the transport characteristics change continuously with sweep-rate, suggesting that ionic migration modifies transport within a pre-existing polarization-controlled electrostatic landscape rather than completely dominating the conduction process. The absence of abrupt switching, together with the robust polarization and electrostatic retention established by PFM and KPFM, indicates that transport in the ultrathin regime is governed by coupled ferroionic dynamics, in which polarization-assisted carrier injection and ionic migration contribute simultaneously to the electrical response. To evaluate the generality of these observations, $I$-$V$ loops were measured and the rectification ratio $R$ was extracted from multiple flakes spanning a broad thickness range (Figure S7). The statistical analysis reveals a systematic evolution from large rectification ratios in ultrathin flakes to weaker rectification in thicker flakes. The sweep-rate-dependent transport measurements clearly reveal a thickness-dependent crossover in the dominant charge-transport mechanism of CCPS.

The distinct transport mechanisms in two thickness regimes become particularly evident from the evolution of the C-AFM hysteresis loops acquired at a slow sweep-rate of 1 V $s^{-1}$ (Figures 4a,b). As discussed earlier, slow voltage sweeps provide sufficient time for $Cu^+$ redistribution and therefore amplify the contribution of ionic motion to charge transport. The ~110 nm flake exhibits a strongly asymmetric hysteresis loop with an abrupt current increase at ~8 V during the forward positive sweep (Figure 4a). Such threshold-like switching is consistent with field-driven $Cu^+$ accumulation beneath the grounded Pt tip, progressively lowering the local injection barrier and promoting the formation of a localized $Cu^+$-rich conductive pathway (Figure 4c). The resulting transition from a high-resistance state (HRS) to a low-resistance state (LRS) resembles the SET[42] process commonly observed in ionic resistive-switching systems[43,44]. During the subsequent reverse sweep, the enhanced conductance persists despite the decreasing bias, indicating that the redistributed $Cu^+$ ions do not immediately relax to equilibrium. Upon polarity reversal, $Cu^+$ ions migrate away from the tip region, increasing the local barrier and disrupting the conductive pathway.

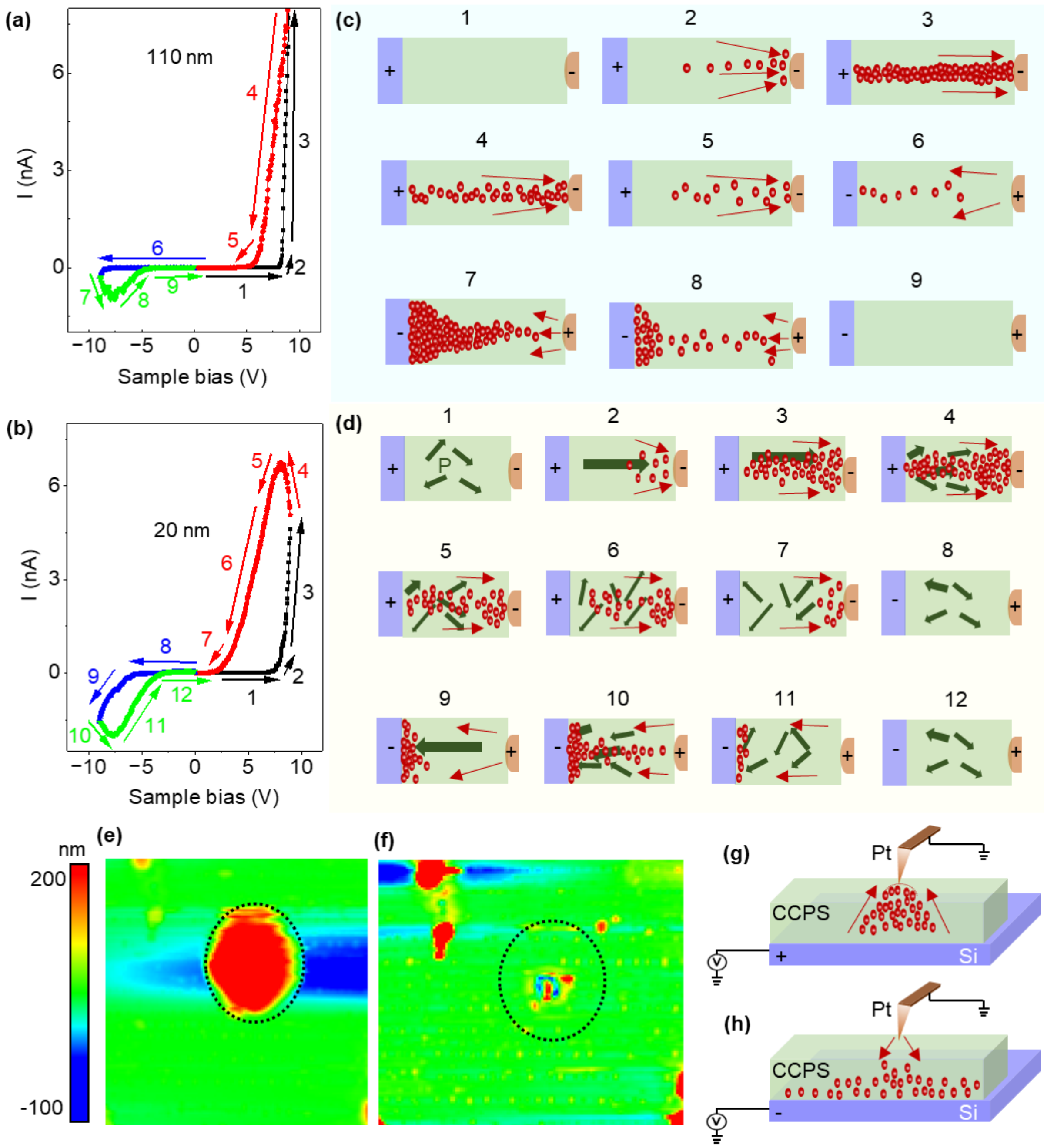

***Figure 4. Distinct switching mechanisms in ultrathin and thicker CCPS flakes.*** *a,b) C-AFM I-V hysteresis loops acquired on representative ~110 nm and ~20 nm CCPS flakes at a voltage sweep-rate of 1 V s*$^{-1}$*. Numbers and colored arrows denote successive stages of the voltage sweep sequence. c,d) Proposed switching mechanisms corresponding to the stages indicated in a) and b), respectively. In the ~110 nm flake (c),* $Cu^{+}$ *ion migration leads to the progressive formation and rupture of conductive ionic filaments. In the ~20 nm flake (d),* $Cu^{+}$ *migration occurs with FE polarization-modulated conduction, resulting in coupled ferroionic transport.*

*Red spheres represent $Cu^+$ ions and green arrows indicate FE polarization direction. e,f) AFM topography images of the ~110 nm flake following five consecutive positive (0 → +9 → 0 V) (e) and negative (0 → −9 → 0 V) (f) voltage sweep semi-cycles. Dashed circles highlight surface protrusions and depressions associated with $Cu^+$-ion accumulation and depletion, respectively. g,h) Schematic illustrations of $Cu^+$-ion accumulation beneath the conductive tip during positive bias (g) and $Cu^+$-ion depletion during negative bias (h), consistent with the observed topographic modifications and ionic-migration-driven switching.*

This process restores the HRS and constitutes the RESET operation. The pronounced hysteresis therefore reflects the strong history dependence associated with ionic redistribution. In contrast, the ~20 nm FE flake displays a qualitatively different response (Figures 4b,d). Rather than exhibiting an abrupt SET transition, the current evolves continuously throughout the voltage cycle. Positive bias simultaneously drives polarization alignment and $Cu^+$ migration toward the tip, while polarization-induced bound charges modify the interfacial electrostatic potential and facilitate carrier injection. The gradual increase in current therefore reflects the coupled evolution of polarization and ionic redistribution. Notably, the current continues to increase even after the applied voltage begins to decrease, indicating that the internal ferroionic state evolves on a longer timescale than the external electrical stimulus. A similar delayed response is observed during the negative voltage cycle, where polarization reversal and ionic migration proceed concurrently. Unlike the ~110 nm flake, where transport is dominated by localized ionic accumulation, the FE state appears to promote a less localized redistribution of $Cu^+$ ions. The resulting hysteresis is therefore consistent with a polarization-assisted ferroionic conduction pathway rather than abrupt ionic-pathway switching. The delayed current maxima observed during both voltage polarities further support the presence of polarization retention as observed in earlier PFM (Figure 1) and KPFM (Figure 2) measurements that are absent in the thicker regime. Direct evidence for substantial ionic redistribution in the 110 nm flake is provided by post-bias topographic imaging. Following five positive voltage semi-cycles (0 V → +9 V → 0 V), a pronounced surface protrusion develops beneath the tip location (Figure 4e), whereas repeated negative semi-cycles generate a localized depression (Figure 4f). These morphological changes are consistent with $Cu^{+}$ accumulation and depletion[23,24], respectively, as illustrated schematically in Figures 4g,h. The asymmetric surface evolution provides independent evidence of electrically driven ionic migration and supports the proposed ionic-pathway formation mechanism responsible for the memristive response of the 110 nm flake. Although direct visualization of conductive pathways is beyond the scope of the present study, the combined evidence from thickness-dependent ferroelectricity, sweep-rate-dependent transport, post-bias topographic evolution, and KPFM measurements consistently supports a crossover from ionic-pathway-dominated conduction in thick flakes to polarization-regulated ferroionic transport in ultrathin FE flakes.

**Nanoscale photoresponse in C-AFM measurements**

Figure 5 presents the evolution of light-modulated charge transport in CCPS, revealing a coupled interplay among photogenerated carriers, FE polarization, and $Cu^{+}$ ionic migration. Light-dependent C-AFM *I–V* hysteresis loops measured under illumination power densities ($\Phi_{ph}$ = 0–77 mW $cm^{-2}$) for representative ~110 nm and ~20 nm flakes at sweep-rates of 20 V $s^{-1}$ and 1 V $s^{-1}$ are shown in Figures 5a–d. In both thickness regimes, increasing $\Phi_{ph}$ leads to a systematic enhancement of current, modification of hysteresis width, and reduction of rectification, indicating strong photoinduced modulation of interfacial charge injection. Under dark conditions, the transport behavior follows the sweep-rate-dependent ferroionic dynamics discussed previously. Since the photon energy used here (2.79 eV for 445 nm wavelength) is larger than the bandgap (1.22 eV)[33] of CCPS, photogenerated electron–hole pairs produced during illumination will increase the free-carrier density in CCPS and partially screen the interfacial Schottky barriers, leading to enhanced conductivity under both bias polarities. At the higher sweep-rate of 20 V $s^{-1}$ (Figures 5a,c), where the contribution of slow ionic motion is relatively suppressed, both 110 nm and 20 nm flakes exhibit similar qualitative trends: increasing $\Phi_{ph}$ enhances current and progressively reduces rectification. This response is therefore dominated by photoelectronic effects, including photocarrier generation and photo-assisted barrier lowering. At the lower sweep-rate of 1 V $s^{-1}$ (Figures 5b,d), where ionic migration contributes more significantly to transport, pronounced differences emerge between the two thickness regimes. For the 110 nm flake (Figure 5b), increasing $\Phi_{ph}$ results in a strong enhancement of current and a widening of the hysteresis loop, particularly under positive bias. At intermediate and high $\Phi_{ph}$, the $I$–$V$ characteristics evolve into strongly asymmetric loops with abrupt current jumps, suggesting a transition toward photo-assisted ionic memristive[45] behavior.

The reduction in onset voltage with increasing $\Phi_{ph}$ (Figure 5e) suggests that optical excitation facilitates interfacial barrier modulation and enhances $Cu^{+}$ mobility, consistent with a combined effect of photocarrier screening and ion-assisted transport. However, the 20 nm FE flake (Figure 5d) exhibits a smoother evolution of the $I$–$V$ loops with $\Phi_{ph}$. Although illumination increases the current and modifies hysteresis, the switching remains continuous without abrupt jump-like transitions, indicating that FE polarization continues to regulate the charge transport even under enhanced ionic activity. The absence of sharp switching suggests a more distributed ferroionic response rather than filament-dominated conduction. To quantify the effect of illumination, the threshold voltages at positive ($V_{th+}$) and negative ($V_{th-}$) voltage sweeps corresponding to the onset of current were extracted from the $I$–$V$ loops (Figure 5e,f). For 110 nm and 20 nm flakes, both $V_{th+}$ and $V_{th-}$ decrease with increasing $\Phi_{ph}$. This pronounced shift reflects enhanced carrier injection facilitated by photocarrier-induced barrier lowering, accompanied by increased ionic contribution under illumination, which together promote easier formation of conductive pathways in thick CCPS flakes.

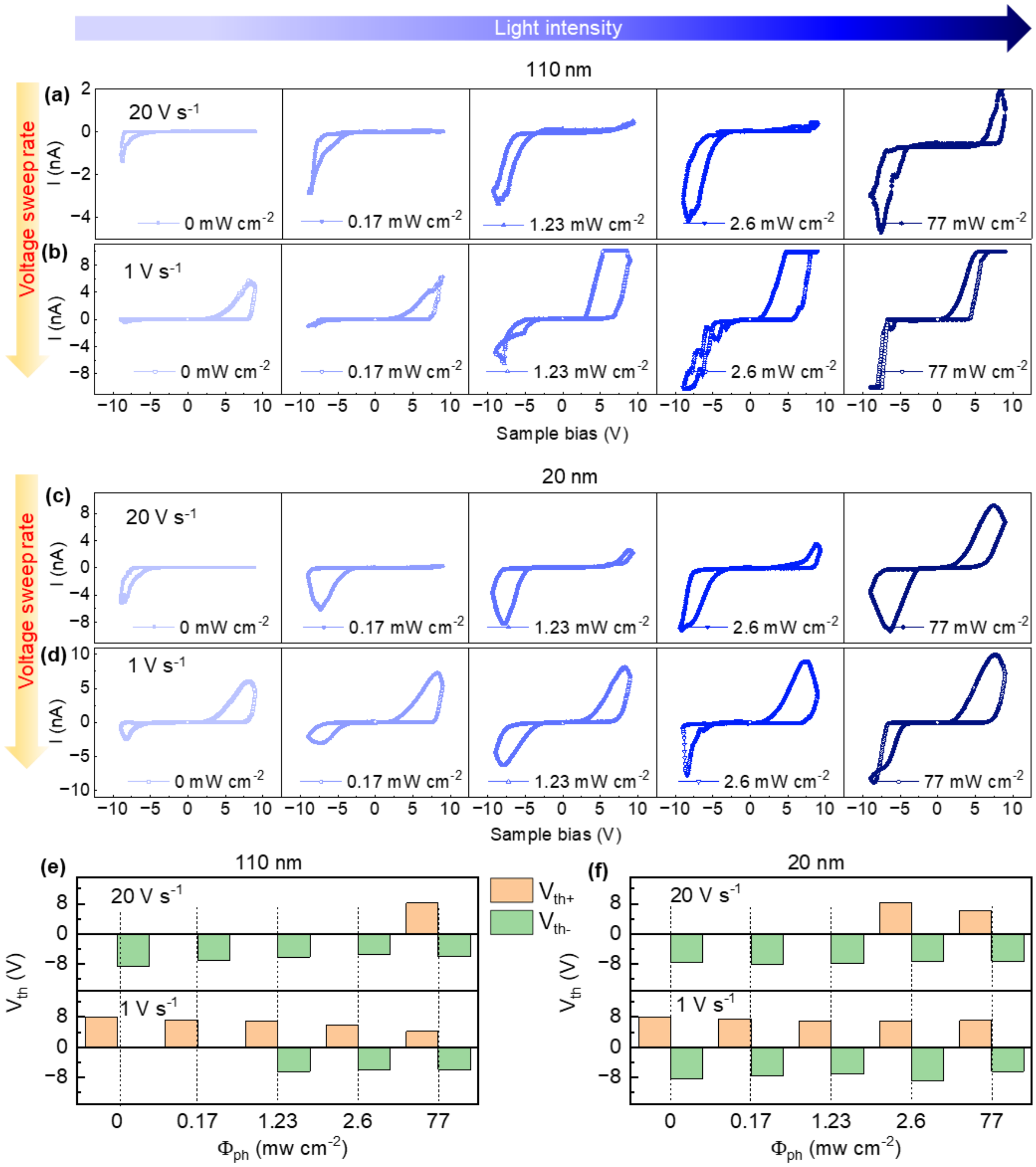

***Figure 5. Illumination and sweep-rate-dependent C-AFM I–V hysteresis in CCPS flakes.*** *a,b) I–V hysteresis loops measured by C-AFM on a ~110 nm CCPS flake under varying* $\Phi_{ph}$ *(0–77 mW cm*$^{-2}$*) at voltage sweep-rates of 20 V s*$^{-1}$ *(a) and 1 V s*$^{-1}$ *(b). c,d) Corresponding C-AFM I–V hysteresis loops for a ~20 nm CCPS flake measured under identical illumination conditions*

*at sweep-rates of 20 V s⁻¹ (c) and 1 V s⁻¹ (d). e,f) Illumination-dependent evolution of the $V_{th}$ extracted from the I–V characteristics for the ~110 nm and ~20 nm flakes, respectively, at sweep-rates of 20 V s⁻¹ (upper panels) and 1 V s⁻¹ (lower panels). Numbers along the x-axis in e,f) represent the value of $\Phi_{ph}$ during measurement, not the axis scale.*

Taken together, the results of Figures 4 and 5 establish a unified picture of thickness-dependent ferroionic transport in CCPS. In thicker flakes, the $Cu^+$ ions redistribute readily under an applied electric field, producing localized $Cu^+$-rich conductive pathways that give rise to memristive-like switching, and prominent rectification changes. Optical excitation further enhances this response by increasing carrier density and reducing interfacial barriers, thereby facilitating ionic redistribution and lowering the threshold for conductive-pathway formation. For ultrathin flakes, robust switchable polarization modifies the local electrostatic landscape and regulates ionic motion. As a result, charge transport evolves through a coupled ferroionic state characterized by gradual hysteresis and current responses rather than abrupt switching events. Illumination strengthens the ferroionic response[36,37,46] by increasing photocarrier density, yet the transport remains continuously modulated by polarization rather than transitioning to filament-dominated conduction. These observations collectively reveal a thickness-controlled crossover from ionic-pathway-dominated memristive-type transport to polarization-regulated ferroionic conduction and demonstrate that FE order fundamentally alters the dynamics of ionic migration in vdW CCPS.

**Device-level measurement of ionic transport**

The results discussed thus far were obtained using scanning probe microscopy (SPM) techniques, where charge transport was probed locally through a conductive AFM tip in a vertical metal/CCPS/substrate geometry. These measurements revealed thickness-dependent ferroionic transport, ranging from polarization-regulated conduction in ultrathin flakes to ionic-pathway-dominated memristive behavior in thicker crystals. To determine whether these local transport characteristics persist at larger length scales, we fabricated two-terminal planar devices using exfoliated CCPS flakes (>100 nm thick) contacted by coplanar Ag electrodes (Figures 6a,b). Figure 6c shows the time-dependent current response measured under different applied biases. At low voltages (< 0.5 V), the current remains nearly constant over time, indicating predominantly electronic conduction. In contrast, higher bias voltages produce a gradual increase in current during electrical stressing, with the magnitude of the current evolution increasing systematically with applied bias.

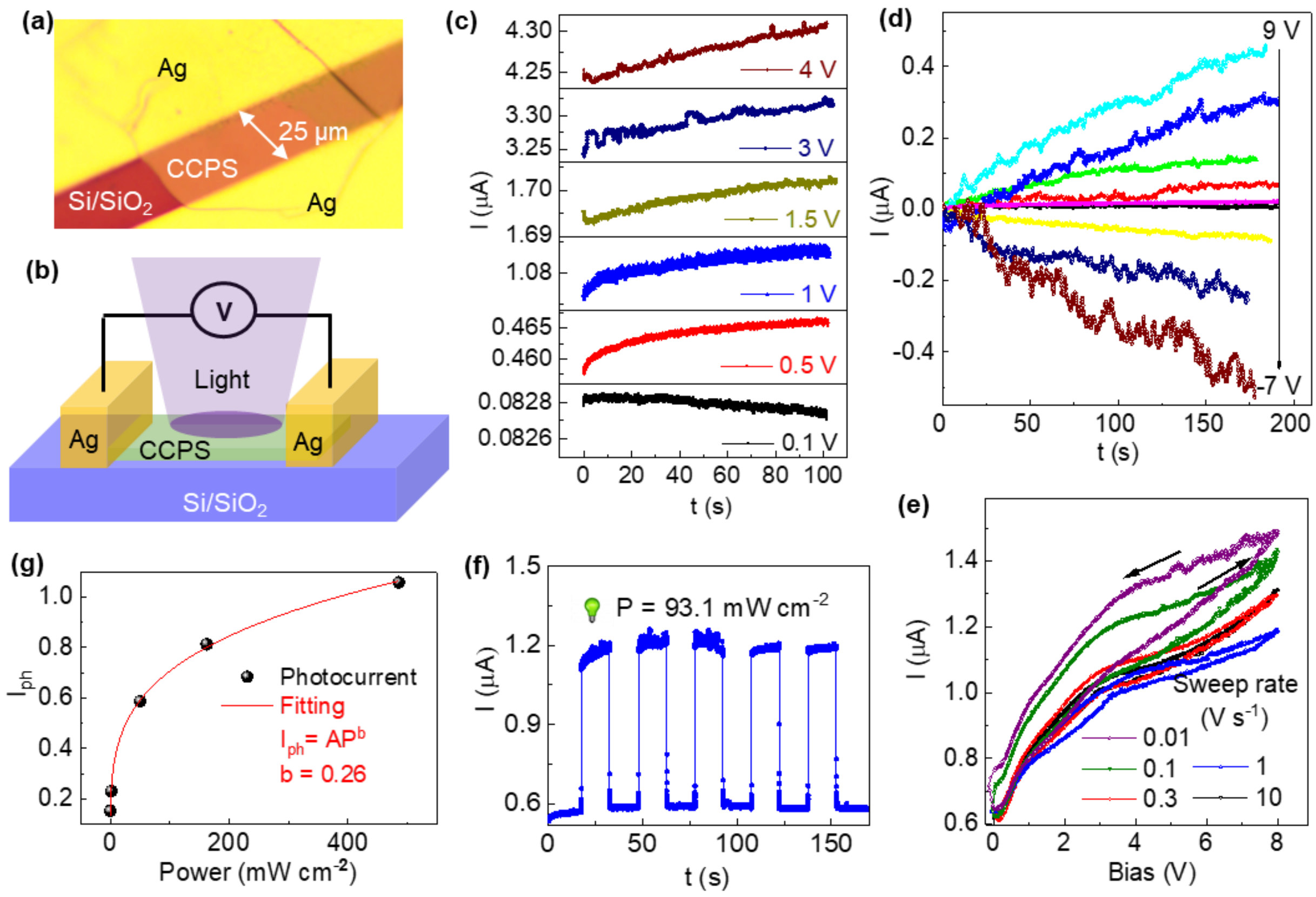

***Figure 6. Device-level ionic and photoresponsive transport in thick CCPS flakes.*** *a,b) Optical microscopy image (a) and schematic representation (b) of a two-terminal CCPS device (>100 nm thick) fabricated on a* $Si/SiO_2$ *substrate with coplanar Ag electrodes. c) Time-dependent current measured under different bias voltages, indicating slow conductance evolution associated with ionic redistribution. d) Current evolution under positive and negative bias*

*polarities over 150 s, highlighting the polarity-dependent nature of ionic transport. e) Dependence of device current (I) on voltage sweep-rate, demonstrating enhanced ionic contribution at slower scan rates. The black arrows indicate the direction of current with voltage sweep. f) Reproducible current modulation under repeated light ON–OFF cycles at a fixed applied bias. g) Photocurrent as a function of illumination power density, showing the power-dependent photoresponse of the device.*

Such behavior is consistent with field-assisted $Cu^+$ redistribution within the CCPS channel, which progressively modifies the conductivity. At 4 V, the pronounced increase in current suggests the emergence of highly conductive ionic pathways, analogous to the ionic transport processes inferred from the nanoscale measurements on thick (~110 nm) CCPS flakes. The conductivity evolution is similar upon reversing the bias polarity (Figure 6d), indicating that the electrically induced ionic redistribution can be reversed within the planar device geometry and directional nature of $Cu^+$ drift under the applied electric field. Further evidence for ionic transport is provided by the sweep-rate-dependence of channel current shown in Figure 6e. Reducing the voltage sweep-rate from 10 V $s^{-1}$ to 0.01 V $s^{-1}$ leads to a substantial increase in current together with a broadening of the hysteresis window. The enhanced current at longer measurement timescales is consistent with the slow ionic dynamics identified in the C-AFM experiments. The optoelectronic response of the device was investigated under 445 nm illumination. Figure 6f shows the current measured during repeated light ON/OFF cycles at a fixed bias of 0.5 V and an illumination power density of ~90 mW $cm^{-2}$. Illumination increases the current from approximately 0.58 μA to 1.18–1.22 μA, and the current returns close to its initial value when the light is removed. The reproducible modulation observed over successive cycles demonstrates stable photoresponse and efficient photocarrier generation in the CCPS channel. The resulting ON/OFF ratio ($I_{light}/I_{dark} \approx 2$) highlights the sensitivity of the device to visible-light excitation. To further quantify the photoresponse, the photocurrent ($I_{ph}$)[47] was analyzed as a function of illumination power density ($P$) (Figure 6g) using the relation $I_{ph} = AP^b$. The extracted exponent of b = 0.26 suggests a strongly sublinear power dependence, consistent with substantial carrier trapping and recombination during phototransport[48,49]. When considered together with the sweep-rate-dependent transport and bias-induced conductivity evolution, this behavior further suggests that slow ionic processes contribute to the overall photoresponse of the device. Clearly, this measurement on a planar device demonstrates that the ferroionic transport identified locally by SPM persist at the device level. The combined observations of bias-dependent changes in conductivity, reversible polarity-dependent switching, sweep-rate-sensitive transport, and illumination-enhanced conduction establish that ionic redistribution remains an important factor governing the electrical and optoelectronic behavior of thick CCPS flakes. These results therefore bridge the nanoscale ferroionic mechanisms revealed by SPM with the operation of practical device architectures and highlight the potential of CCPS for photoresponsive memristive and neuromorphic technologies.

## 3. Conclusion

In summary, we have systematically investigated the interplay between ferroelectricity, ionic transport, and photoresponse in vdW CCPS across a broad thickness range using complementary nanoscale and device-level measurements. We find that room-temperature ferroelectricity emerges in ultrathin flakes below ~40 nm, whereas thicker crystals exhibit suppressed switchable polarization and transport dominated by $Cu^+$ ionic migration. This thickness-dependent evolution produces a crossover from polarization-regulated ferroionic transport in ultrathin flakes to ionic-pathway-dominated memristive conduction in thicker crystals. The C-AFM measurements reveal fundamentally different transport dynamics in these two thickness regimes. Thick flakes exhibit conductance evolution consistent with field-driven ionic redistribution, while ultrathin flakes display smooth hysteretic transport governed by the coupled evolution of polarization and ionic motion. Optical excitation further modifies these dynamics by increasing carrier density and reducing interfacial barriers, resulting in photo-assisted ionic transport in thicker flakes and enhanced ferroionic responses in ultrathin samples. Device-level measurements demonstrate that the nanoscale transport phenomena persist in planar architectures, linking local ferroionic processes to macroscopic electrical and optoelectronic functionality. These results establish FE order as a key parameter governing ionic transport in CCPS and reveal dimensionality as an effective route for tuning the balance between polarization and ion migration in vdW ferroionic materials. More broadly, our findings provide a framework for engineering coupled photoferroionic phenomena and highlight CCPS as a promising platform for multifunctional memory, sensing, and neuromorphic device technologies.

## 4. Methods

### CCPS crystal growth, exfoliation, and device fabrication

CCPS single crystals were synthesized by the chemical vapor transport (CVT) method as described previously[18]. The phase purity and composition of the as-grown crystals were verified by X-ray diffraction[18] (XRD), Raman spectroscopy, and energy-dispersive X-ray spectroscopy (EDS). Thin CCPS flakes were mechanically exfoliated from bulk crystals using adhesive tape (Nitto SPV224 PVC) and transferred onto highly doped Si and Pt-coated Si substrates. Flake thicknesses were identified using optical microscopy and AFM. For SPM measurements, the conductive substrate served as the bottom electrode, while the conductive AFM probe acted as the top electrode. Two-terminal planar devices were fabricated by transferring CCPS flakes onto Si substrates coated with a 300 nm thick $SiO_2$ layer. Approximately 50 nm thick Ag electrodes were deposited by thermal evaporation through a shadow mask to define a channel length of 25 μm. Electrical transport measurements were performed using a dual-channel source meter (Keithley 2602B), which provided simultaneous voltage sourcing and current measurement.

**SPM measurements**

AFM, PFM, KPFM, and C-AFM measurements were carried out at room temperature under ambient conditions (~25 °C and 20–30% relative humidity) using a Park Systems NX10 atomic force microscope. OOP PFM measurements were performed using Pt-coated conductive probes with an AC excitation voltage of 1 V superimposed on a DC bias sweep of ±10 V. Local polarization switching experiments were conducted using a box-in-box poling protocol. KPFM measurements were acquired in amplitude-modulated mode using Au-coated conductive probes to map the surface potential and $V_{CPD}$. For C-AFM measurements, Pt-coated conductive probes with a nominal force constant of 18 N $m^{-1}$ were employed. The electrical bias was applied through the bottom electrode while the conductive probe was maintained at ground potential.

**Optical measurements**

Optoelectronic measurements were performed using a continuous-wave diode laser ($\lambda$ = 445 nm) with a beam diameter of ≈ 2 mm. The illumination intensity was controlled using the laser drive current and calibrated neutral-density filters. For both SPM and device-level measurements, the laser beam uniformly illuminated the measurement region. Control experiments performed on bare substrates under identical illumination conditions showed negligible changes in surface potential or electrical transport, confirming that the observed photoresponses originated from the CCPS flakes rather than substrate-related or measurement-induced artifacts.

## Supporting Information

This Supporting Information includes Raman and EDS analyses for structural and compositional verification, along with AFM, PFM, and C-AFM measurements supporting thickness-dependent ferroelectric and transport behavior in CCPS nanoflakes.

## Acknowledgements

This work has been funded by the Department of Defense through Grant No. W - 911NF2120213. Part of this work was also supported by the Center for Energy Efficient Magnonics, an Energy Frontier Research Center funded by the U.S. Department of Energy, Office of Science, Basic Energy Sciences, under Contract No. DE-AC02-76SF00515.

## Author contributions

S.C. performed the experiments, carried out the investigation and methodology, analyzed the data, and wrote the original draft of the manuscript. A.B. contributed to the two-terminal device measurements. R.B. grew the single crystals. S.C. and R.C.B. jointly planned the experiments and the latter also supervised the project, acquired funding and resources, and reviewed and edited the manuscript. All authors discussed the results and approved the final version of the manuscript.

## Conflict of Interest

The authors declare no conflicts of interest.

## Data availability

The data that support the findings of this study are available from the corresponding author upon request.

## Table of Contents

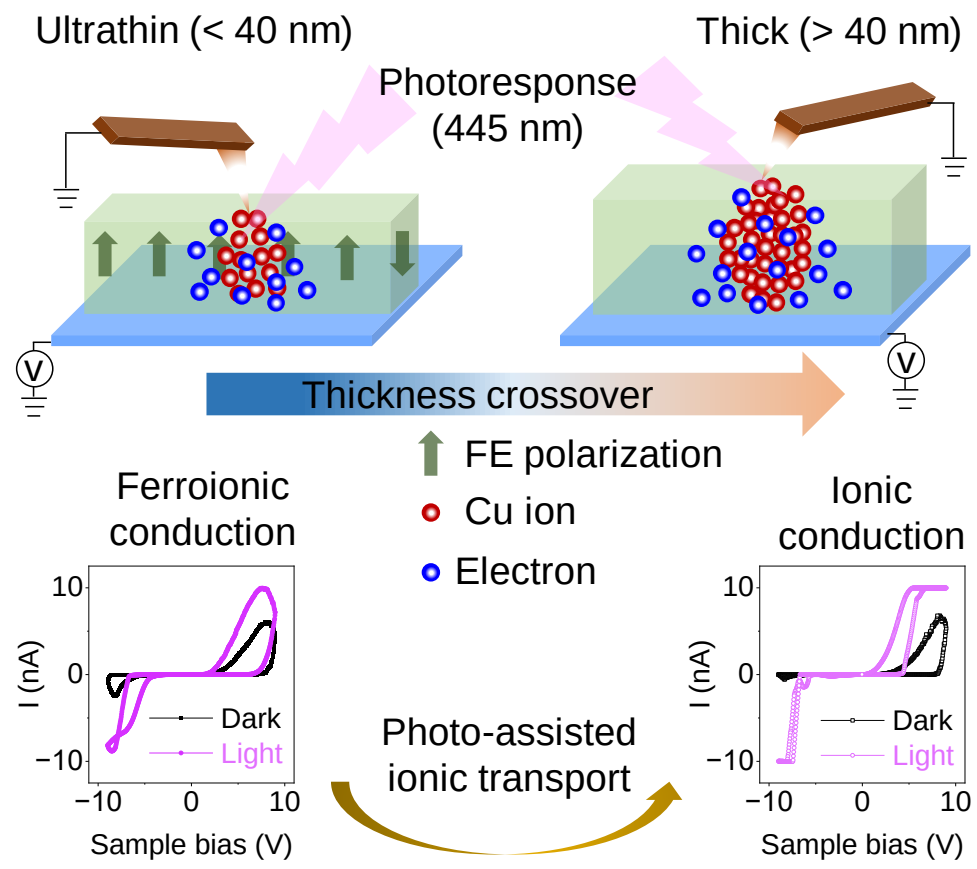

Thickness-dependent modulation of $CuCrP_2S_6$ governs the competition between ferroelectric polarization and $Cu^+$ ion migration, producing a crossover from polarization-regulated ferroionic transport in ultrathin flakes to ionic-pathway-dominated memristive transport in thicker crystals. Optical illumination further enhances these coupled ferroionic processes, enabling tunable photoresponsive transport.

# Supplementary Material

# Thickness-Controlled Ferroelectric-to-Ionic Transport in $CuCrP_2S_6$ at Room Temperature

Subhashree Chatterjee, Abhishek Bajgain, Rabindra Basnet, Ramesh C. Budhani

Department of Physics, Morgan State University, Baltimore, MD, 21251, USA

E-mail: ramesh.budhani@morgan.edu

This Supplementary Information provides additional experimental details and supporting data for the study entitled "Thickness-Controlled Ferroelectric-to-Ionic Transport in $CuCrP_2S_6$ (CCPS) at Room Temperature." The supplementary figures include Raman and EDS analyses for phase and compositional verification, as well as AFM, PFM, and C-AFM measurements supporting the thickness-dependent evolution of ferroelectricity and charge transport in exfoliated crystals of CCPS. This additional information provides further insight into the crossover from ferroelectric-controlled transport to ion-mediated conduction with the increasing thickness of crystals. These results complement and support the conclusions presented in the main manuscript.

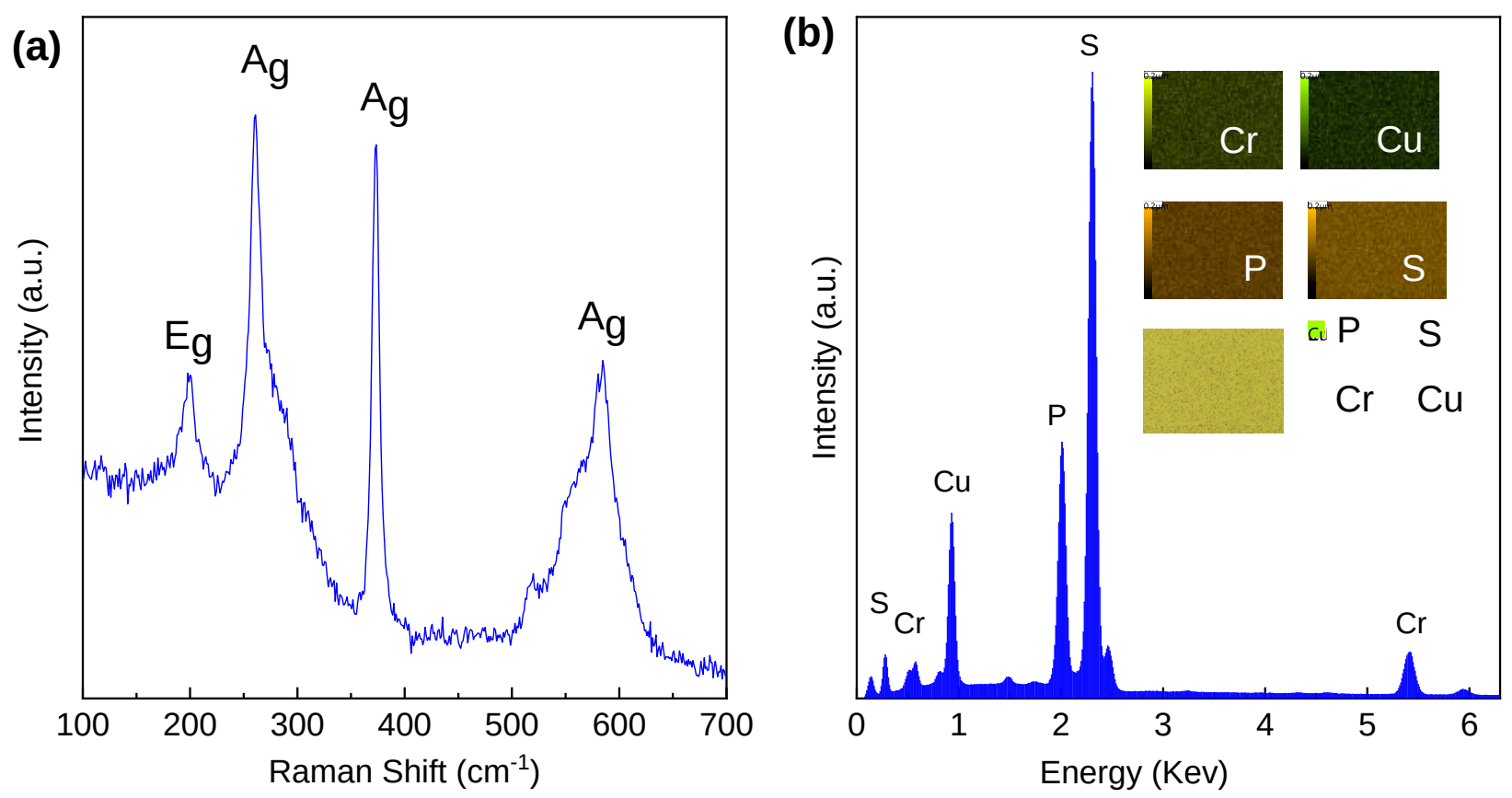

***Figure S1.*** *Room temperature a) Raman, and b) EDS X-ray spectrum of a CCPS single crystal. Raman vibration modes are consistent with earlier reported literatures.* [1] *EDS spectrum of the CVT-grown CCPS single crystal showing individual peaks for each element. The inset displays EDS elemental mapping of Cr, Cu, P, and S over a larger area of the bulk crystal, confirming the uniform distribution of all four elements through the volume. The single crystal phase, crystallinity and composition indicate good crystal growth.*

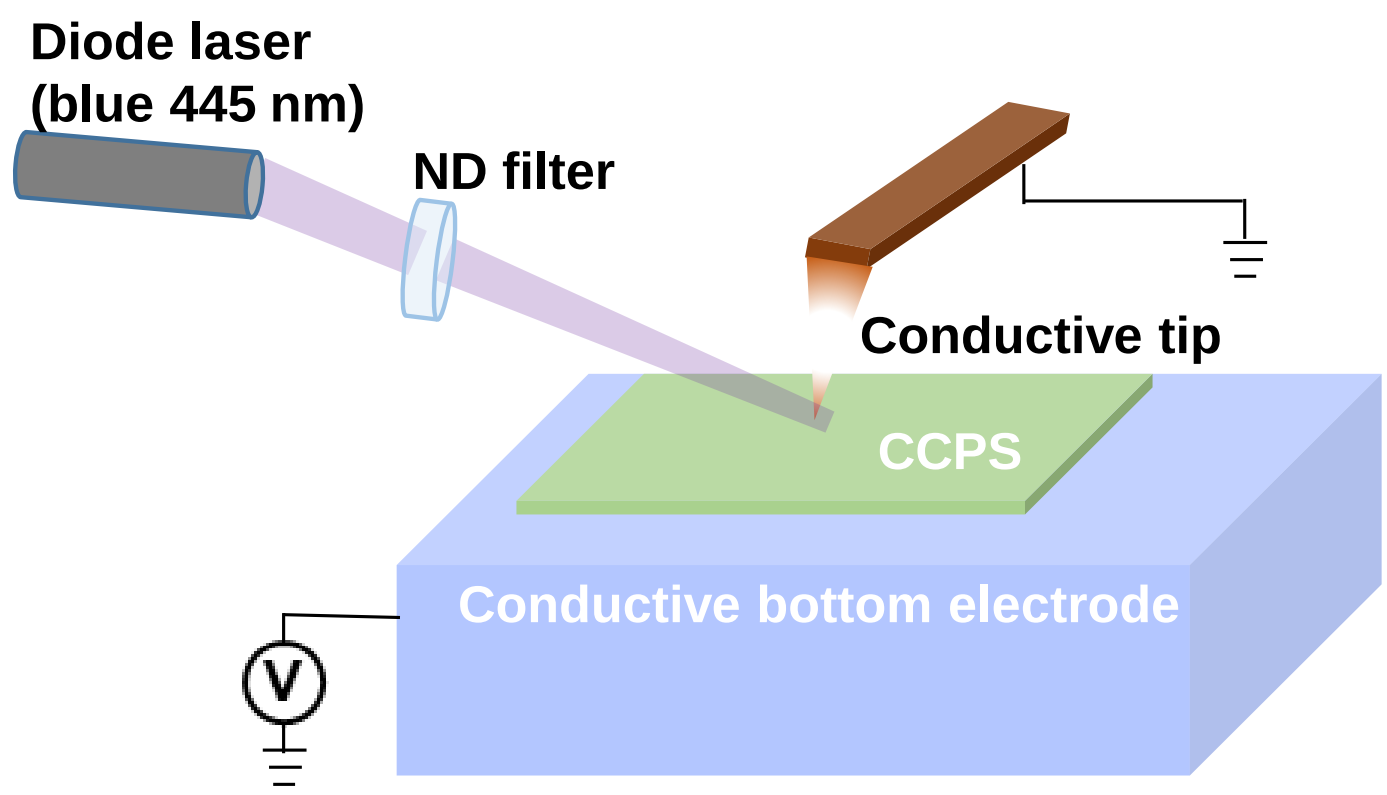

***Figure S2.*** *Nanoscale SPM measurement configuration set up under controlled optical illumination, as discussed in the Methods section.*

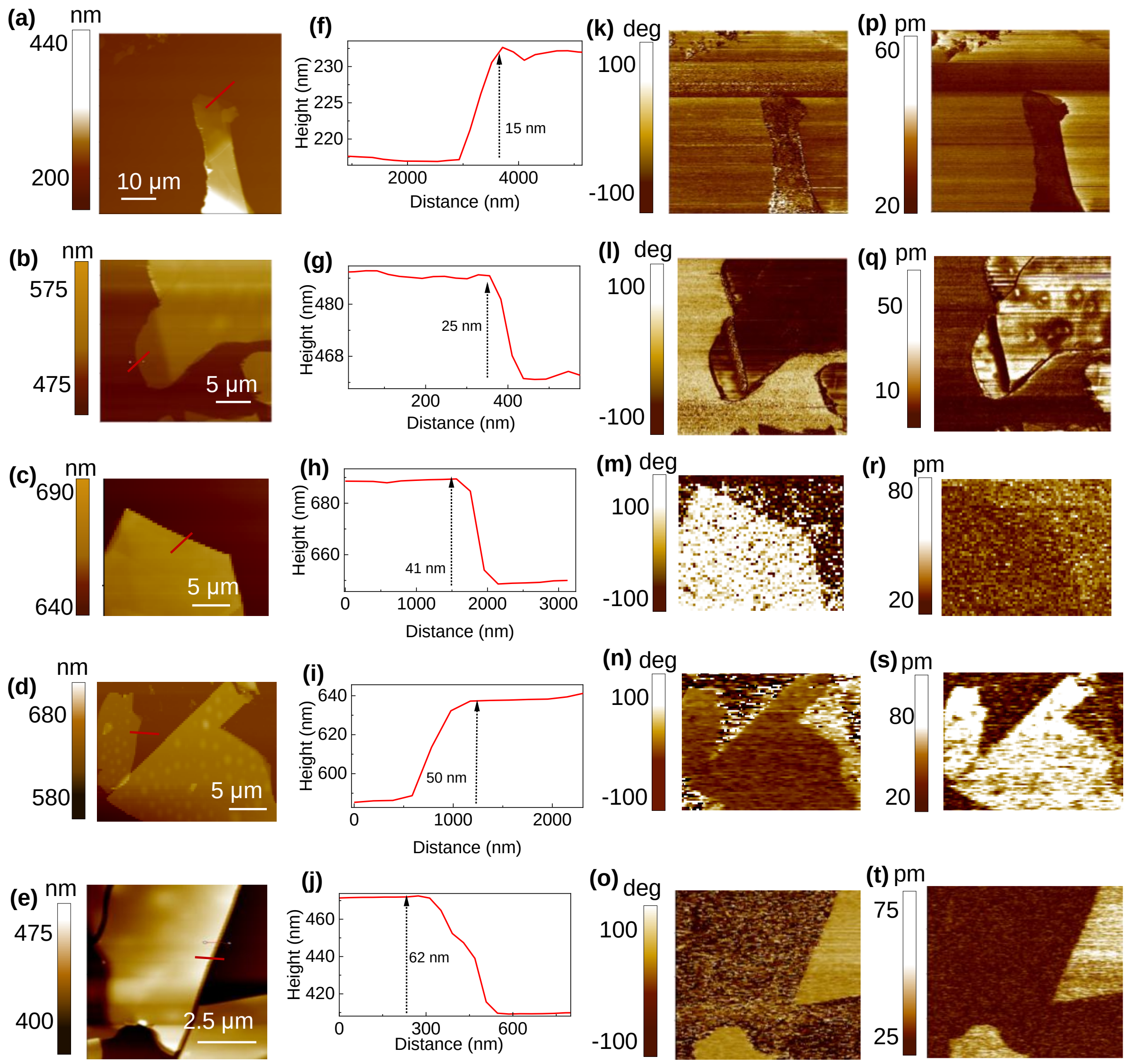

***Figure S3.*** *a–e) AFM surface topography images of exfoliated CCPS flakes with thicknesses ranging from ~15 to 62 nm. f–j). Corresponding height profiles extracted along the red lines indicated in panels (a–e), confirming the thickness of each flake. k–o) Out-of-plane (OOP) PFM phase images of the same flakes. p–t) Corresponding PFM amplitude images, highlighting the electromechanical response across the flakes.*

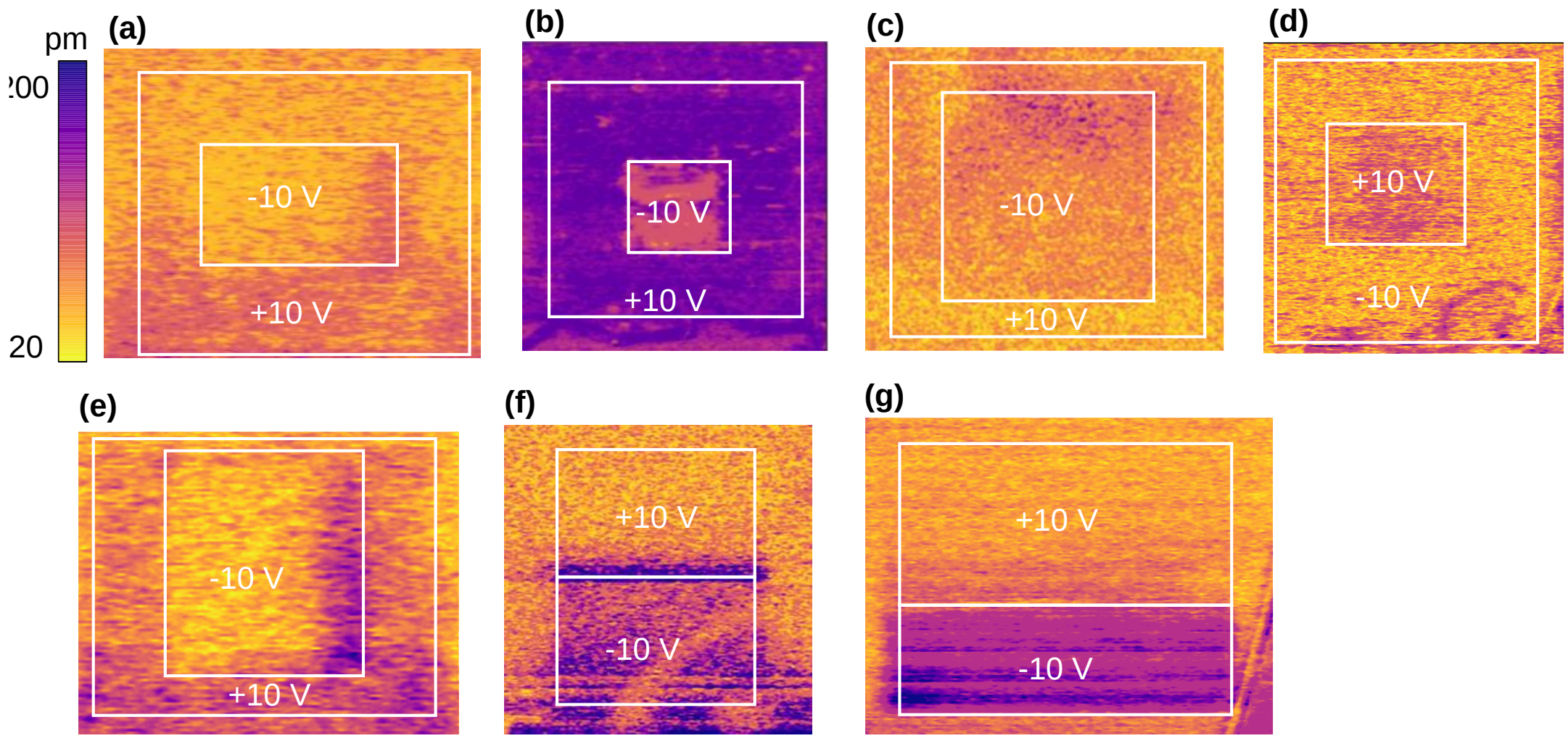

***Figure S4.*** *a–g) OOP-PFM phase images of CCPS flakes with thicknesses of (a) 12 nm, (b) 17 nm, (c) 20 nm, (d) 30 nm, (e) 38 nm, (f) 50 nm, and (g) 62 nm, respectively, after local domain writing using reversible DC poling with ±10 V bias.*

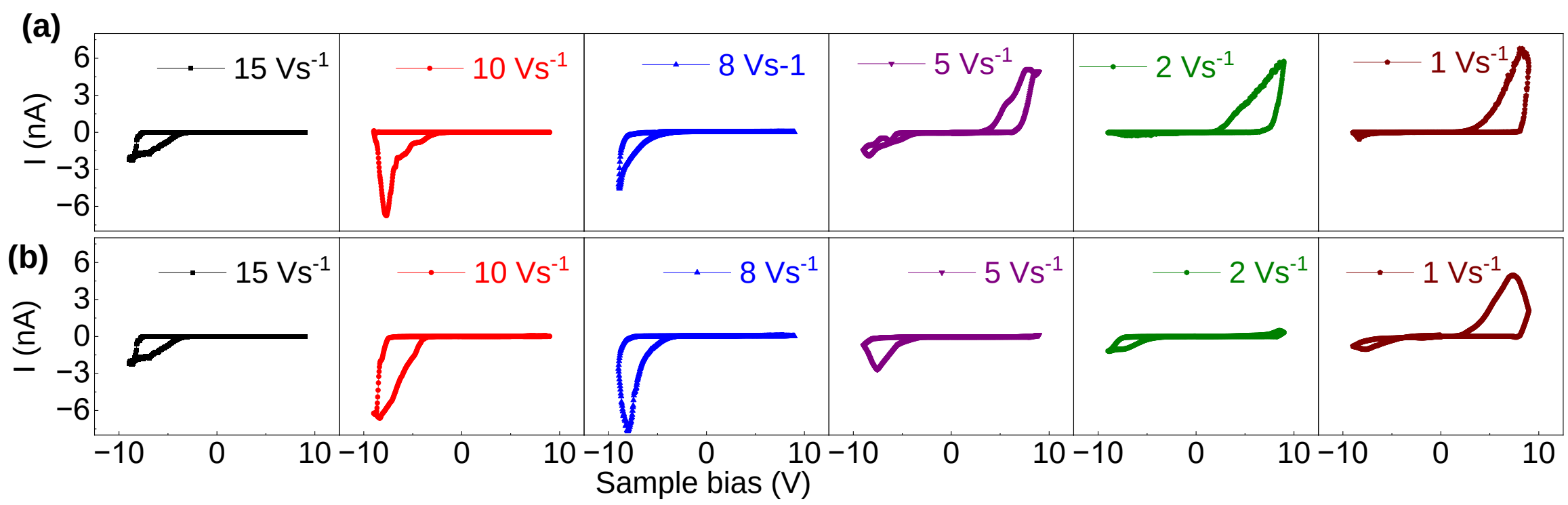

***Figure S5.*** *Response of current–voltage (I–V) curves to sweeping speed with a triangular voltage sweeping full-cycle of (0 V→+9 V→-9 V→0 V), for a) 110 nm and b) 20 nm flakes, respectively.*

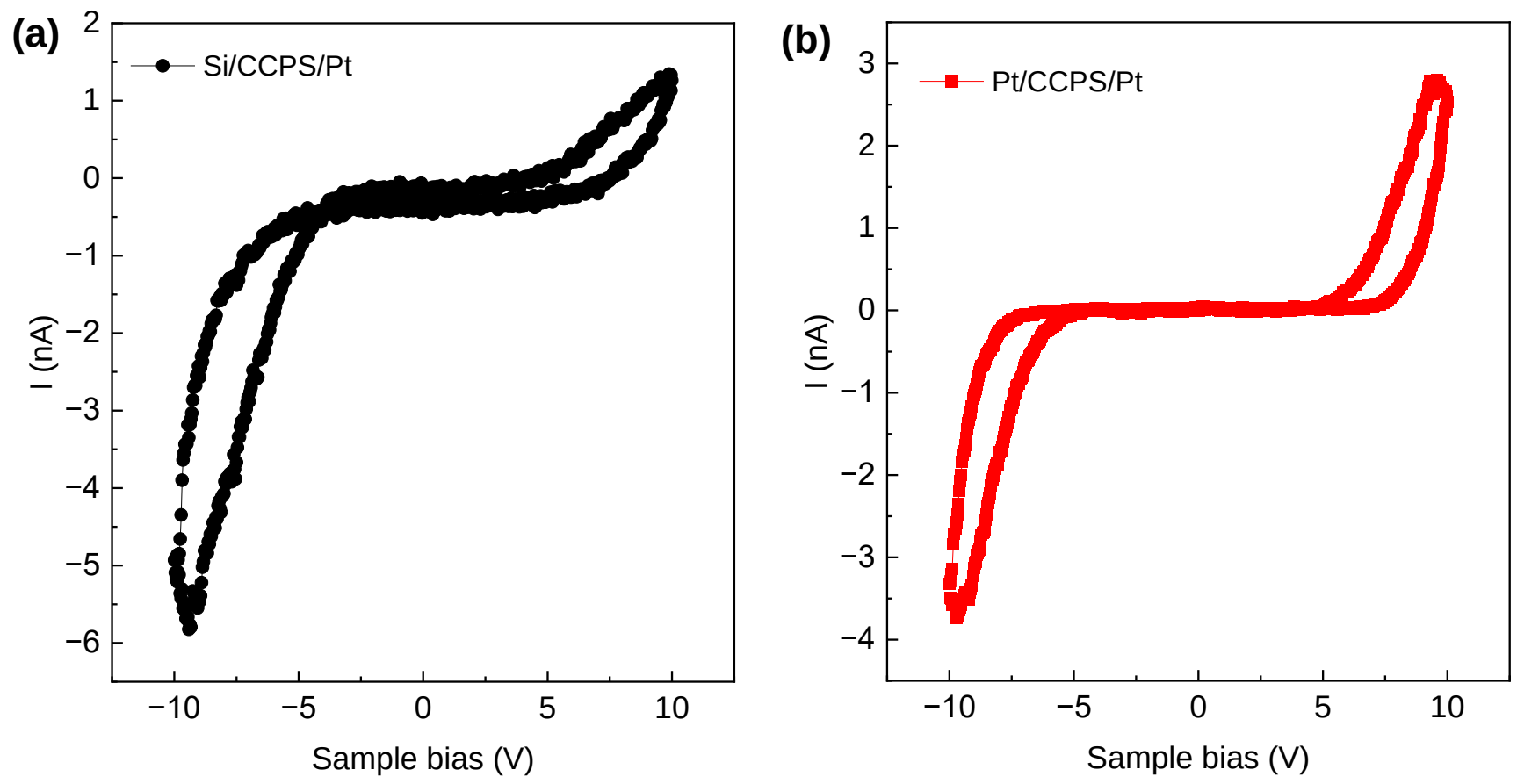

***Figure S6.*** *a,b) Room temperature I-V loops of CCPS measured in asymmetric Si/CCPS/Pt-tip configuration and symmetric Pt/CCPS/Pt-tip configuration, respectively. In two cases, CCPS is exfoliated on conductive Si and Pt coated Si used as the bottom electrode during C-AFM measurements using Pt tip as top electrode.*

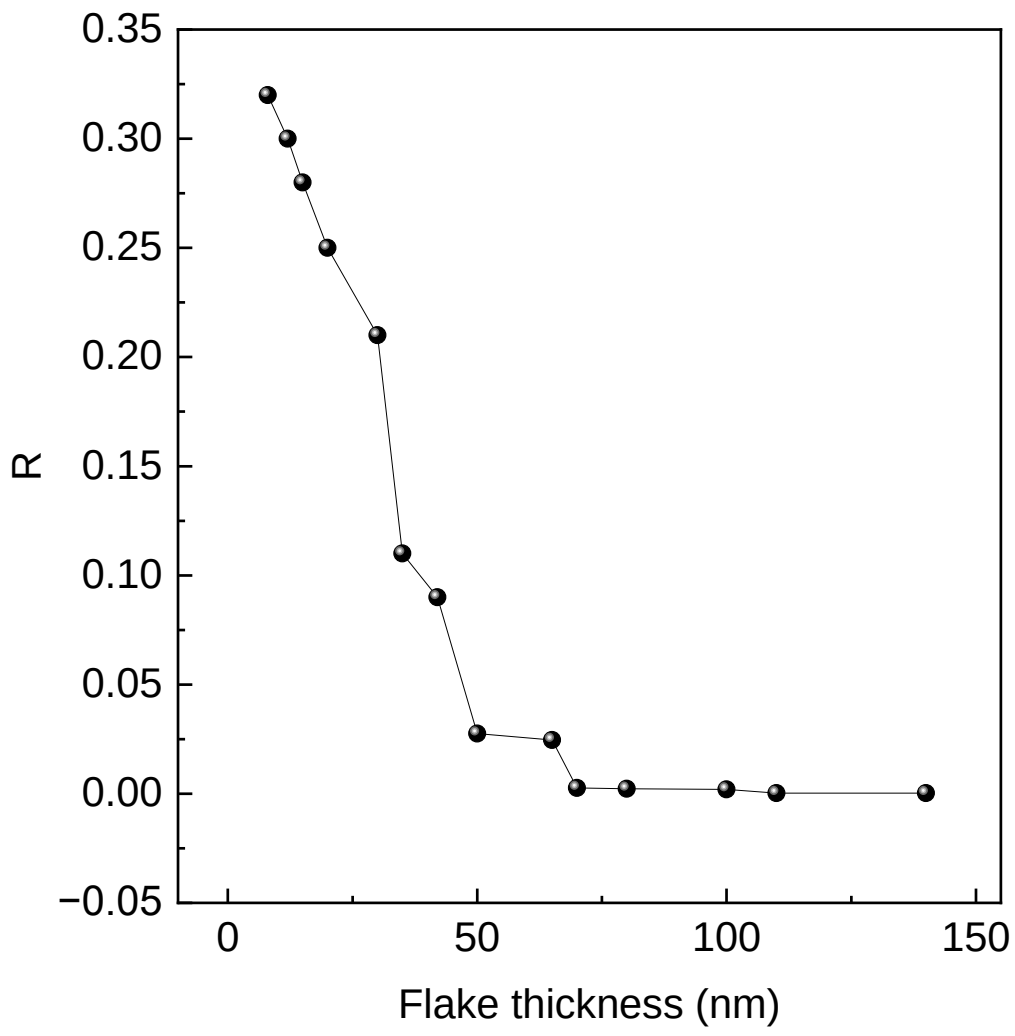

***Figure S7.*** *Rectification ratio $R = |I_{-}/I_{+}|$ calculated for different CCPS flake thickness for a voltage sweep rate 1V $s^{-1}$.*